\documentclass{cernrep} 
\usepackage{texnames}
\usepackage[T1]{fontenc}
\usepackage[bookmarks, colorlinks=true, linktoc=page, linkcolor=black, citecolor=black, urlcolor=black]{hyperref}
\usepackage[strings]{underscore}    
\newcommand\dottcmd[1]{\texttt{#1}\endgroup}
\newcommand{\vn}{\begingroup\catcode`\_=11 \catcode`\%=11 \dottcmd}

\usepackage{fancyhdr}

\fancyhfoffset{4 mm}
\fancypagestyle{ARTTITLE}{%
\fancyhf{} 
\lhead{\small{Proceedings of the 2018 CERN--Accelerator--School course on\\
\it{Numerical Methods for Analysis, Design and Modelling of Particle Accelerators}, Thessaloniki, (Greece)}} 
\lfoot{Available online at \url{https://cas.web.cern.ch/previous-schools}}
\rfoot{\thepage\hspace*{3mm}}
 
}

\begin{document}

\title{Simple Linear and Nonlinear Examples of Truncated Power Series Algebra}
 
\author {\'Etienne Forest}

\institute{KEK, Tsukuba, Japan}

\begin{abstract}
The paper displays calculations of linear systems as explained by Dr.~Guido Sterbini. We also show a simple nonlinear calculation involving a rotation followed by an octupole kick.
Some analytical calculations are compared to the~Truncated Power Series Algebra (TPSA) results. The examples use the library PTC which is in MAD-X of CERN and BMAD of Cornell (see Ref.~\cite{bmada}).
\end{abstract}


\maketitle 
 
\thispagestyle{ARTTITLE}
 

\section{Introduction and recommended reference}
 
My work is documented in at least two books Refs. \cite{thebook, thebook2}. In this article I will show simple linear and nonlinear examples  of Truncated Power Series Algebra (TPSA) usage within the code PTC.
PTC  is inside the code  MAD-X (CERN's code) and  BMAD (Cornell's code). For completeness, the main programs and the MAD-X file are included in the text in Sections   \ref{sec:zrotationoctupole}, \ref{sec:zguidolecture} and \ref{sec:madfiles}.

The reader unwilling to compile these programs can at least look at the software to get an idea of what a TPSA-based perturbation tool kit entails.

The actual software can be found at
 \url{http://www.takafumi.org/cas/cas_paper/}. There are three archives: one for the code FPP/PTC, one for the MAD-X file and the flatfile for PTC, and finally the~two main programs used in this report. 
 The MAD-X file is not guaranteed to work at this point in time with the version of PTC inside MAD-X at CERN so you might elect to simply run PTC  as a standalone program guided by  the included ``Makefiles.''
 
 I advice the reader to look into the CAS  2015 lectures of Dr. Werner Herr delivered in Warsaw, Poland. They go into some of the same material in more detail without reference to a specific code. 
 
 \subsection{Needed software tools}\label{sec:soft}
 
 Three things are needed: a tracking code, a TPSA package  and an analysis package.  The tracking code we will use, called PTC, has various features that are irrelevant to the present paper. The most ``unusual'' feature of PTC is to separate the beam line from the sequence of magnets. The beam line in PTC is not a sequence of magnets but a sequence of pointers to a magnet. This allows for the natural creation of shared magnets as they exist in colliders, recirculators and ``figure 8'' geometries. That sequence is made of a Fortran90 type called \vn{fibre}. However for this paper we can simply ignore the difference between the fibre and the magnet. 
 
 If you do not understand what this is all about, please simply think that fibre equals magnet: you will not miss the central message of this paper. 
 
 So in summary we have conceptually three distinct  pieces of software:
 
 \begin{enumerate}
 
 \item We need a package that can turn the  tracking of floating numbers into the tracking of Taylor series.  The package I used will be referred to as ``FPP'' which stands for ``fully polymorphic package'', a name coined by Frank Schmidt of CERN. It is based on Martin Berz's original ``DA'' package developed  at LBNL as explained in Refs. \cite{berz,tpsa}. Its main types are called ``polymorphs'' because they can change at execution time from real to Taylor series. 
 
 \item We  need a tracking code which can track a ray. This is the code PTC. This code is equipped with FPP. Therefore, rays can be real or Taylor series because the program PTC is coded with the~polymorphs of FPP.
 
 \item Finally we need the ability to analyse Taylor maps. This is located in  a routine of FPP called Ci_tpsa.f90. It is linked to a complex version of the LBNL version of Berz's package. It manipulates Taylor maps and analyses them. It is a nonlinear extension of a typical matrix analysis tool: it can diagonalise the linear part of a  map and extends its analysis  to the  nonlinear case. Basically it performs canonical perturbation theory on one-turn maps rather than on the equations of motion which are too messy and complex in a realistic accelerator. This was the central message of my first book in  \cite{thebook} but  it is demonstrated in great details in \cite{thebook2}.
 
 \end{enumerate}
 
  \subsection{Notational conventions for Fortran types in the text}\label{sec:conv}

 \begin{enumerate}
 
 \item The types belonging to the tracking code, which may or may not contain  polymorphs, will be written in lower case. For example, the type \vn{fibre} which contains magnets is written in lower case. The real polymorph themselves belong to the Fortran90 type \vn{real_8}. 
 
 \item The types belonging to analysis will be denoted  by capital letters. For example, the type for a map made of Taylor series will be denoted by \vn{C_DAMAP}.
 
  \end{enumerate}
 
  One should note that Fortran90 does not make a distinction between lower case and capitals. This is simply a convention adopted in the text of this paper so as to avoid constantly pointing out what is from the tracking code and what is from the analysis package.

 \subsection{A Tracking code is needed}\label{sec:coded}

A tracking code is a program which allows the computation of a final position in terms of an initial position in phase space. Phase space for the examples of this paper will be: 
%
\begin{equation} \vec{r}=
\left({x,{p}_{x},y,{p}_{y},\delta ,\tau }\right). \label{eq:r}\end{equation}
 In some cases, when it makes sense, it might be only 
%
\begin{equation} \vec{r}=
\left({x,{p}_{x},y,{p}_{y}}\right), \label{eq:r4}\end{equation}
or just
%
\begin{equation} \vec{r}=
\left({x,{p}_{x}}\right). \label{eq:r2}\end{equation}

The variables  $x$ and $y$ of  \Eref{eq:r} represent respectively the transverse horizontal and vertical position.  The variable $p_x$ and $p_y$ are the transverse  momenta associated to the positions $x$ and $y$. In a drift space where  particles propagate freely, these momenta are related to the variables
%
\begin{equation} {x}^{\prime }=
{dx \over dz}=
{{p}_{x} \over \sqrt {{\left({1+\delta }\right)}^{2}-{p}_{x}^{2}-{p}_{y}^{2}}}~~~{\rm a}{\rm n}{\rm d} ~~~{y}^{\prime }=
{dy \over dz}=
{{p}_{y} \over \sqrt {{\left({1+\delta }\right)}^{2}-{p}_{x}^{2}-{p}_{y}^{2}}}, \label{eq:mom}\end{equation}
where the variable $z$ is the direction of propagation perpendicular to $x$ and $y$.  The variable $\delta $ is the relative momentum variable:
%
\begin{equation} \delta =
{p-{p}_{0} \over {p}_{0}}. \label{eq:del}\end{equation}
The variable $\tau $ is a time-like variable.

We will not waste time discussing the details of the tracking code except to say that it produces a map consistent with the Hamiltonian nature of the flow. Sometimes, the small angle approximation is used in accelerator physics. In that case \Eref{eq:mom} reads
%
\begin{equation} {x}^{\prime }=
{dx \over dz}=
{{p}_{x} \over 1+\delta }~~~{\rm a}{\rm n}{\rm d} ~~~{y}^{\prime }=
{dy \over dz}=
{{p}_{y} \over 1+\delta }. \label{eq:moms}\end{equation}

In the variables of  \Eref{eq:r}, the Hamiltonian nature of the flow is equivalent to the statement that the flow is symplectic: a word you will hear over and over again in accelerator physics.

\subsection{Some details of the code}\label{sec:code}

The code used in this paper is called ``PTC.''  In this code, particles are tracked from ``fibres'' to ``fibres.'' For the purpose of this paper, we can confuse the fibre with the magnet. So here is a typical call from the~code PTC:

\begin{verbatim}

call propagate(cell,xs,+state,fibre1=i,fibre2=j)

\end{verbatim}

\vn{cell} is a layout, which amounts to a beam line in all our examples.  \vn{xs} is a  Fortran90 object of type  ``\vn{probe}''. It contains the ray and more things such as spin.  The ray part is contained in xs\%x(1:6). \vn{state} is an object of type \vn{internal_state}: it controls certain settings of the  tracking code particularly with respect to Taylor series. For example, the  \vn{internal_state} ``\vn{only_4d}'' forces the Taylor series to depend on the 2-d-f phase space of \Eref{eq:r4}.

If ``fibre2'' is omitted, then the call to \vn{propagate} tracks a particle from position ``i'' back to position ``i.'' Obviously we are dealing with the one-turn map  of a ring.

\begin{verbatim}

call propagate(cell,xs,+state,fibre1=i)

\end{verbatim}

Repetitive calls to \vn{propagate} allows a user to produce a phase plot in the variables of \Eref{eq:r}.

\subsection{Producing a Taylor map}\label{sec:maketaylor}
 
Let us look at the following code fragment:

\begin{verbatim}
closed_orbit(1:6)=0.d0;                                      
call find_orbit_x(cell,closed_orbit(1:6),STATE,1.e-8_dp,fibre1=1)  
                  .
                  .
                  .
xs0=closed_orbit(1:6)   ! xs0 contains orbit and spin 
id=1   !    identity map
xs=id+xs0   !  xs is a probe_8 which can become a Taylor series
 
call propagate(cell,xs,+state,fibre1=1)  ! computes one turn map around closed orbit
 
one_turn_map=xs
\end{verbatim}

This fragment is part of the program z_guido_lecture.f90, displayed in Appendix \ref{sec:zguidolecture}, which we will use to reproduce some of the results of Dr.~Guido Sterbini's lecture. Here are the objects involved:

\begin{enumerate}
\item  \vn{closed_orbit(1:6)} is a array of six real numbers which contains the closed orbit at position 1 in the ring.
\item  \vn{xs0} is an object of type \vn{probe}. It is the ray in its full glory. For example it contains spin as well as the six variables of \Eref{eq:r}.
\item \vn{id} is a Taylor map which is set to identity; it is of type \vn{C_DAMAP}.
\item  \vn{xs} is a ray of type \vn{probe_8}. This ray has the ability to turn itself into Taylor series. 
\item  The assignment \vn{xs=id+xs0 } takes   the closed orbit (in \vn{xs0}) adds it to the identity Taylor map and assigns it to the the ray \vn{xs}. 
\item \vn{ propagate(cell,xs,+state,fibre1=1)} tracks a ray from position 1 back to position 1. 
\item The routine propagate returns the final value of \vn{xs} after one-turn. This ray contains presumably a~Taylor series. A type \vn{C_DAMAP} is created by the assignment \vn{one_turn_map=xs}.

\end{enumerate}

\subsection{What does ray and the  Taylor map look like?}\label{sec:xsmap}

Let us start with the ray \vn{xs} of type \vn{probe_8}. We did a mid-plane symmetric calculation. In that case the description of the dynamics can be restricted to one degree of freedom (1-d-f), i.e., \Eref{eq:r2}. The~code PTC track up to six phase space variables, spin and other things. In the mid-plane symmetric case  below, variables 1 and 2 depend quadratically on the first and second variables, namely $x$ and $p_x$.  So for example,
%
\begin{equation} {x}^{f}=
\ 0.3643\ x\ +\ 10.37\ {p}_{x}\ +\ 16.99\ {x}^{2}-108.5\ x{p}_{x}-411.3\ {p}_{x}^{2} \label{eq:ex}\end{equation}
is, truncated to 4 digits, what is represented by  ``\vn{Variable~~~1}'' just below.

The third, fourth and fifth ($\delta $) variables are simply zero on the closed orbit. The sixth variable, which is the path length, depends on the the 1-d-f phase of \Eref{eq:r2}. In this particular machine, one can see that the 1-d-f system is totally self-consistent.

\begin{verbatim}
  The Ray of type Probe_8
  ORBIT 
  Variable            1

 Properties, NO =    2, NV =    2, INA =   19
 *********************************************

   1  0.3643938681973571       1  0
   1   10.37284171884971       0  1
   2   16.99977446004454       2  0
   2  -108.5189415529075       1  1
   2  -411.3420203011964       0  2
  Variable            2

 Properties, NO =    2, NV =    2, INA =   42
 *********************************************

   1 -0.8331176555867445E-01   1  0
   1  0.3727292207573401       0  1
   2  0.2310072333707236       2  0
   2   3.185240693854261       1  1
   2   194.8817617298971       0  2
  Variable            3

 Properties, NO =    2, NV =    2, INA =   62
 *********************************************

           0  Real Polynomial is zero 
  Variable            4

 Properties, NO =    2, NV =    2, INA =   57
 *********************************************

           0  Real Polynomial is zero 
  Variable            5
  0.000000000000000E+000
  Variable            6

 Properties, NO =    2, NV =    2, INA =   56
 *********************************************

   1 -0.1561490325699064E-05   1  0
   1 -0.1106422811325558E-04   0  1
   2  0.1425247852379894E-01   2  0
   2   2.534406506937603       1  1
   2  -9.317500948809176       0  2
  SPIN X 
        .
        .
        .
\end{verbatim}

The map \vn{one_turn_map} of type \vn{C_DAMAP} is extracted from    ray \vn{xs} of type \vn{probe_8}.  Maps of type \vn{C_DAMAP} can become complex during analysis. Therefore, each coefficient is made of two real numbers. Here  is the printout. Notice that our map is real and thus all the imaginary parts are zero:

\begin{verbatim}
  The Taylor Map of type C_DAMAP
   
           2  Dimensional map 

          1, NO =    2, NV =    2, INA =  155
 *********************************************

    I  COEFFICIENT          ORDER   EXPONENTS
      NO =     2      NV =     2
   1  0.3643938681973571       0.000000000000000       1  0
   1   10.37284171884971       0.000000000000000       0  1
   2   16.99977446004454       0.000000000000000       2  0
   2  -108.5189415529075       0.000000000000000       1  1
   2  -411.3420203011964       0.000000000000000       0  2
    -5   0.000000000000000       0.000000000000000       0  0

          1, NO =    2, NV =    2, INA =  154
 *********************************************

    I  COEFFICIENT          ORDER   EXPONENTS
      NO =     2      NV =     2
   1 -0.8331176555867445E-01   0.000000000000000       1  0
   1  0.3727292207573401       0.000000000000000       0  1
   2  0.2310072333707236       0.000000000000000       2  0
   2   3.185240693854261       0.000000000000000       1  1
   2   194.8817617298971       0.000000000000000       0  2
    -5   0.000000000000000       0.000000000000000       0  0
  Spin Matrix is identity 
  c_quaternion is identity 
 No Stochastic Radiation 
\end{verbatim}

One notices that the Taylor map contains only the 1-d-f part: The final values of $x$ and $p_x$ are in terms of their initial values via the interpretation of  \Eref{eq:ex}.
 
\section{A linear example: z_guido_lecture.f90}

In this section, I reproduce some of Dr.~Sterbini's results using TPSA and a file produced by the code PTC embedded inside the CERN code MAD-X.

 The MAD-X lattice file is displayed in section \ref{sec:madfile}. 
The MAD-X code invokes a ``PTC script'' through a file in Section \ref{sec:scriptfile} whose main function is to produce a flat file input for PTC.

The PTC program is in Section \ref{sec:zguidolecture} and it reads the flat file produced by MAD-X.

\subsection{The input file}

\begin{verbatim}
type(layout),pointer :: cell
     .
     .
     .
!!!! reading the flat file produced by MAD-X
call ptc_ini_no_append
call read_lattice_append(M_U,"../../MAD-X/files_for_cas/als_guido/one_cell.txt")
cell=>m_u%start
\end{verbatim}

The file \vn{one_cell.txt} is a so-called flat file describing the lattice magnet by magnet: it is produced by PTC via the code MAD-X. It is loaded into a linked list of type \vn{mad_universe} and it is of type \vn{layout}. The Fortran90 pointer \vn{cell} is of type \vn{layout} and allows us to give a convenient name to the~layout under study through the assignment \vn{cell=>m_u

So we now have a lattice stored into the layout ``\vn{cell}.'' The layout is a linked list of fibres, i.e., our magnets. We are ready to do some calculations.

\subsection{Normalising of the Taylor map}\label{sec:ana}

 The map having been produced as described in \Sref{sec:maketaylor}, we must analyse it. This is done by the~following call:
 
 \begin{verbatim}
 
 call c_normal(one_turn_map,normal_form,phase=phase_one_turn_map)  

  \end{verbatim}

The object  \vn{normal_form} is of type \vn{C_NORMAL_FORM} and  \vn{phase(1:3)} is an array of three Taylor series of type \vn{C_TAYLOR}. The type \vn{C_NORMAL_FORM}  is defined as
{
\small
 \begin{verbatim}

TYPE C_NORMAL_FORM
TYPE(C_DAMAP) A1 ! BRINGS TO FIX POINT AT LEAST LINEAR
TYPE(C_DAMAP) A2 ! LINEAR NORMAL FORM 
TYPE(C_FACTORED_LIE) G ! NONLINEAR PART OF A IN PHASORS
TYPE(C_FACTORED_LIE) KER ! KERNEL I.E. NORMAL FORM IN PHASORS
TYPE(C_DAMAP) A_T ! TRANSFORMATION A (M=A N A^-1) 
TYPE(C_DAMAP) N !TRANSFORMATION N (M=A N A^-1) 
TYPE(C_DAMAP) AS ! FOR SPIN (M = AS A N A^-1 AS^-1) 
TYPE(C_DAMAP) ATOT ! FOR SPIN (M = ATOT N ATOT^-1) 
! STORES RESONANCES TO BE LEFT IN THE MAP, INCLUDING SPIN (MS)
INTEGER NRES,M(NDIM2T/2,NRESO),MS(NRESO) 
REAL(DP) TUNE(NDIM2T/2),DAMPING(NDIM2T/2),SPIN_TUNE !@1 STORES SIMPLE INFORMATION
LOGICAL POSITIVE ! FORCES POSITIVE TUNES (CLOSE TO 1 IF <0)
!!!ENVELOPE RADIATION STUFF TO NORMALISE RADIATION (SAND'S LIKE THEORY)
COMPLEX(DP) S_IJ0(6,6) ! EQUILIBRIUM BEAM SIZES
COMPLEX(DP) S_IJR(6,6) ! EQUILIBRIUM BEAM SIZES IN RESONANCE BASIS
 ! EQUILIBRIUM EMITTANCES AS DEFINED BY CHAO (COMPUTED FROM S_IJR(2*I-1,2*I) I=1,2,3 )
REAL(DP) EMITTANCE(3)
END TYPE C_NORMAL_FORM
  \end{verbatim}
 }
 Mathematically, we have
%
\begin{equation} one\_turn\_map=
normal\_form\%atot\circ normal\_form\%n\circ {normal\_form\%atot}^{-1}.\label{eq:norm}\end{equation}
The Fortran90 code corresponding to \Eref{eq:norm} is

 \begin{verbatim}
 one_turn_map=normal_form%atot*normal_form%n*normal_form%atot**(-1)
  \end{verbatim}
 
The operation  \vn{*} between two Taylor maps (type \vn{C_DAMAP}), performs a map concatenation where the constant part of the map, if present, is ignored. In other words, all these maps are assumed to be around the closed orbit. They are no ``feed down'' effects from a finite closed orbit.

In the linear example of this section, the map \vn{normal_form

\subsection{Freedom in choosing a canonical transformation}\label{sec:cs}

The map $M$ can be normalised with a canonical transformation $A$, which in the linear 1-d-f, amounts to a unit determinant matrix:
%
\begin{equation} M=
A\circ R\circ {A}^{-1}. \label{eq:acs}\end{equation}
In \Eref{eq:acs} the concatenation symbol $\circ $ is left to emphasize that the maps can be nonlinear, i.e., they are not necessarily matrices.  We can modify \Eref{eq:acs} by inserting a rotation $\rho $ since it commutes with $R$:
%
\begin{equation} M=
A\circ \rho \circ R\circ {\rho }^{-1}\circ {A}^{-1}=
B\circ R\circ {B}^{-1}~~{\rm w}{\rm h}{\rm e}{\rm r}{\rm e}~~B=
A\circ \rho .\label{eq:acsr}\end{equation}
Looking at \Eref{eq:acsr}, it is clear that if $\rho $ is a rotation, then both $A$ and $B$ normalise the map $M$. Therefore, we have an infinite number of choices for the map $A$.  But first, let us get the invariant of the map $M$ in terms of $A$.
We start by noticing that a rotation such as $R$ admits the radius as an invariant. Therefore, we can write:
%
\[ {\rm i}{\rm f}~~~{r}^{2}=
{x}^{2}+{p}_{x}^{2}=
\left({{x}^{2}+{p}_{x}^{2}}\right)\circ R, \]
%
\begin{equation} {\rm t}{\rm h}{\rm e}{\rm n}~~~\left({{x}^{2}+{p}_{x}^{2}}\right)\circ {A}^{-1}=
\left({{x}^{2}+{p}_{x}^{2}}\right)\circ {A}^{-1}\circ \underbrace{A\circ R\circ {A}^{-1}}\limits_{M}^{}.\label{eq:csa} \end{equation}
We then conclude that the Courant-Snyder invariant is:
%
\begin{equation} \varepsilon =
\left({{x}^{2}+{p}_{x}^{2}}\right)\circ {A}^{-1}=
\gamma {x}^{2}+2\alpha x{p}_{x}+\beta {p}_{x}^{2}.\label{eq:csaf} \end{equation}
In terms if $A^{-1}$, we can derive the following formulae from \Eref{eq:csaf}:
%
\[ \gamma =
{\left({{A}_{11}^{-1}}\right)}^{2}+{\left({{A}_{21}^{-1}}\right)}^{2}, \]
%
\[ \beta =
{\left({{A}_{12}^{-1}}\right)}^{2}+{\left({{A}_{22}^{-1}}\right)}^{2}, \]
%
%
\begin{equation} \alpha =
{A}_{11}^{-1}{A}_{12}^{-1}+{A}_{21}^{-1}{A}_{22}^{-1}.\label{eq:latfi} \end{equation}
It is easy to show, using the fact that $A$ has a unit determinant, that the lattice functions obey the relation:
%
\begin{equation} 1+{\alpha }^{2}=
\beta \gamma . \label{1abg}\end{equation}

\subsubsection{de Moivre's form for the matrix $M$}\label{sec:demoivre}

Using the symplectic condition\footnote{In the 1-d-f case, one can also used the fact that $A$ has a unit determinant.},
%
%
\begin{equation} AS{A}^{{\intercal }}=
S~~~~{\rm w}{\rm h}{\rm e}{\rm r}{\rm e}~~~~S=
\left({\begin{array}{cc}0&1\\
-1&0\end{array}}\right),\label{eq:sympcond} \end{equation}
 we can express $A$ in terms of $A^{-1}$ 
%
\begin{equation} A=
\left({\begin{array}{cc}{A}_{22}^{-1}&-{A}_{12}^{-1}\\
-{A}_{21}^{-1}&{A}_{11}^{-1}\end{array}}\right), \label{eq:aai}\end{equation}
and using the lattice functions of \Eref{1abg} we get the de Moivre form for $M$:
%
\begin{equation} M=
\left({\begin{array}{cc}\cos\left({\mu }\right)+\alpha \sin\left({\mu }\right)&\beta \sin\left({\mu }\right)\\
-\gamma \sin\left({\mu }\right)&\cos\left({\mu }\right)-\alpha \sin\left({\mu }\right)\end{array}}\right). \label{eq:demoivre}\end{equation}

\subsubsection{The Courant-Snyder choice}\label{sec:cs:ch}
%
\begin{equation} A=
\left({\begin{array}{cc}\sqrt {\beta }&0\\
-\alpha /\sqrt {\beta }&1/\sqrt {\beta }\end{array}}\right)~~~{\rm a}{\rm n}{\rm d} ~~~{A}^{-1}=
\left({\begin{array}{cc}1/\sqrt {\beta }&0\\
\alpha /\sqrt {\beta }&\sqrt {\beta }\end{array}}\right).\label{eq:csc}\end{equation}

This is the choice preferred in accelerator physics. It has several advantages:

\begin{enumerate}
\item Experimentally it induces a phase advance which corresponds to the phase difference between two beam position monitors.
\item\label{it:mat} Mathematically it simplifies perturbation theory because our perturbing potentials depend in the~lowest order on positions only.
\item For example, it follows from item \ref{it:mat} that two thin sextupoles $\pi$ apart in phase advance exactly cancel for a properly chosen strength. 
\end{enumerate}

\subsubsection{The Anti-Courant-Snyder choice}\label{sec:acs:ch}

This  transformation can be guessed by symmetry:

%
\begin{equation} B=
\left({\begin{array}{cc}1/\sqrt {\gamma }&-\alpha /\sqrt {\gamma }\\
0&\sqrt {\gamma }\end{array}}\right)~~~{\rm a}{\rm n}{\rm d} ~~~{B}^{-1}=
\left({\begin{array}{cc}\sqrt {\gamma }&\alpha /\sqrt {\gamma }\\
0&1/\sqrt {\gamma }\end{array}}\right).\label{eq:csca}\end{equation}

\subsubsection{Canonization: choosing $A$}\label{sec:can}

The transformation  \vn{normal_form
This is achieved by a call to the subroutine  \vn{c_canonise(normal_form

\begin{verbatim} 
courant_snyder_teng_edwards=.true.
                     .
                     .
                     .
call c_canonise(normal_form%atot,a_cs) 
\end{verbatim}

The difference between \vn{normal_form
%
\begin{equation} {A}^{-1}B=
\left({\begin{array}{cc}1/\sqrt {\beta }&0\\
\alpha /\sqrt {\beta }&\sqrt {\beta }\end{array}}\right)\left({\begin{array}{cc}1/\sqrt {\gamma }&-\alpha /\sqrt {\gamma }\\
0&\sqrt {\gamma }\end{array}}\right)=
{1 \over \sqrt {1+{\alpha }^{2}}}\left({\begin{array}{cc}1&-\alpha \\
\alpha &1\end{array}}\right).\label{eq:aib}\end{equation}

\subsubsection{The general numerical theory: finding an invariant}\label{sec:vec}

The general problem of perturbation theory is the search for invariants.  The equation that follows applies to nonlinear maps as well:
%
\begin{equation} \varepsilon (x)=
\varepsilon \left({M(x)}\right).\ \ \label{eq:inv}\end{equation}
Symbolically, \Eref{eq:inv} can be written using the concatenation symbol:
%
\begin{equation} \varepsilon =
\varepsilon \circ M.\ \ \ \ \label{eq:invc}\end{equation}
To solve \Eref{eq:invc}, we study the effect of the map $M$ on linear functions since we can build arbitrary functions in terms of linear functions through a Taylor series. Consider an arbitrary linear function $f$:
%
\begin{equation} \ f\left({\vec{r}}\right)=
{v}_{1}{r}_{1}+{v}_{2}{r}_{2}\ \ \ \ \label{eq:linf}\end{equation}
For example, if we have two functions $f$ and $g$, then:
%
%
\begin{equation} \ \left({fg}\right)\circ M=
\left({f\circ M}\right)\ \left({g\circ M}\right).\ \ \ \label{eq:compo}\end{equation}
Let us apply \Eref{eq:invc} to \Eref{eq:linf} :
%
\[ f\circ M\ \ \Rightarrow \ f\left({M\vec{r}}\right)=
\sum\limits_{i=
1,2}^{} {v}_{1}{M}_{1i}{r}_{i}+{v}_{2}{M}_{2i}{r}_{i}\ \ \ \ \  \]
%
\[ \Downarrow  \]
%
\begin{equation}   f\circ M=
{w}_{1}{r}_{1}+{w}_{2}{r}_{1}\Rightarrow \ \vec{w}=
{M}^{{\intercal }}\vec{v}.\ \label{eq:mct} \end{equation}
Let us assume that we diagonalized ${M}^{{\intercal }}$ using some numerical method package:
%
\begin{equation} {M}^{{\intercal }}=
\Gamma \Lambda {\Gamma }^{-1}.\ \ \ \label{eq:diagmt} \end{equation}
The matrix $\Lambda $ is diagonal and contains, in the stable case, the following eigenvalues:
%
\begin{equation} {\Lambda }_{11}=
\exp\left({-i\mu }\right)~~~{\rm a}{\rm n}{\rm d} ~~~{\Lambda }_{22}=
\exp\left({i\mu }\right).\ \label{eq:lam} \end{equation}
Then the eigenvectors ${\vec{\zeta }}_{\mp}$ can be built as follows:
%
\begin{equation} {\vec{\zeta }}_{\mp}=
\Gamma {\vec{\xi }}_{\mp}~~~~{\rm w}{\rm h}{\rm e}{\rm r}{\rm e}~~~~~{\vec{\xi }}_{-}=
\left({\begin{array}{c}1\\
0\end{array}}\right)~~{\rm a}{\rm n}{\rm d} ~~{\vec{\xi }}_{+}=
\left({\begin{array}{c}0\\
1\end{array}}\right).\label{eq:phasv} \end{equation}
We define the eigenfunctions from the above eigenvectors:
%
\begin{equation} {f}_{\mp}(\vec{r})=
{\zeta }_{\mp 1 }{r}_{1}+{\zeta }_{\mp 2 }{r}_{2}\ .\label{eq:eigf}\end{equation}
It follows that:
%
\begin{equation} {\cal M}{f}_{\mp}=
{f}_{\mp}\circ M=
\exp\left({\mp i\mu }\right){f}_{\mp}.\ \label{eq:eigmu} \end{equation}

The notation ${\cal M}$ is used by Dragt to denote the composition map $\circ M$ which acts on function.  Using \Eref{eq:eigmu}, we can easily construct an invariant:%
%
\begin{equation} {\cal M}{f}_{+}{f}_{-}=
{f}_{+}\circ M\ {f}_{-}\circ M\ =
\ \exp\left({i\mu }\right){f}_{+}\exp\left({-i\mu }\right){f}_{-}=
{f}_{+}{f}_{-}.\label{eq:invff} \end{equation}

\subsubsection{ Imposing the Hamiltonian structure on $\Gamma $ of  \protect\Eref{eq:diagmt}}\label{sec:vecs}

The transformation $\Gamma $ is complex and not related to a symplectic matrix, i.e., to a matrix with unit determinant in 1-d-f. This amounts to a normalization of ${\vec{\zeta }}_{\mp}$. Since ${\vec{\zeta }}_{+}$ and ${\vec{\zeta }}_{-}$ are complex conjugate, let us pick  ${\vec{\zeta }}_{-}$. The eigenfunction $f_{-}$ can be written as:
%
\begin{equation} {f}_{-}(\vec{r})=
\underbrace{\left({{\zeta }_{- 1 }^{r}{r}_{1}+{\zeta }_{- 2 }^{r}{r}_{2}}\right)}\limits_{{x}^{n}}^{}\ +\ i\ \underbrace{\left({{\zeta }_{- 1 }^{i}{r}_{1}+{\zeta }_{- 2 }^{i}{r}_{2}}\right)}\limits_{{p}_{x}^{n}}^{}.\label{eq:eigfri}\end{equation}

In \Eref{eq:eigfri}, we assume that the complex eigenvector   $\vec{\zeta}_{-}$ has been normalized so that the Poisson bracket of the real part $x^n$ with the complex part $p_x^n$ is one:
%
%
\begin{equation}  \left[{{x}^{n},{p}_{x}^{n}}\right]=
{\zeta }_{- 1 }^{r}{\zeta }_{- 2 }^{i}-{\zeta }_{- 2 }^{r}{\zeta }_{- 1 }^{i}=
1. \label{eq:pbc}\end{equation}
One notices that the invariant defined as $\varepsilon = f_{+} f_{-}$ is the radius 
%
${{x}^{n}}^{2}+{{p}_{x}^{n}}^{2}$
and therefore the variables $(x^n,p_x^n)$ move on a circle. Therefore, we can construct $A^{-1}$ of \Eref{eq:acs} using \Eref{eq:eigfri}:
%
\begin{equation} {A}^{-1}=
\left({\begin{array}{cc}{\zeta }_{- 1 }^{r}&{\zeta }_{- 2 }^{r}\\
{\zeta }_{- 1 }^{i}&{\zeta }_{- 2 }^{i}\end{array}}\right). \label{eq:ai}\end{equation}
As for ${\Gamma }_s$, the symplectic version of  ${\Gamma }$, it is given by:
%
\begin{equation} {\Gamma }_{s}=
{{A}^{-1}}^{{\intercal }}{{C}^{-1}}^{{\intercal }},\label{eq:gams}\end{equation}
where ${{C}^{-1}}$ is given by \Eref{eq:ci}.

\subsubsection{ Trying on an example}\label{sec:aexp}

Let us do a simple example. For example, if we have 
%
\begin{equation} M=
\left({\begin{array}{cc}1&1\\
-1&0\end{array}}\right). \label{eq:me}\end{equation}
We now diagonalize $M^\intercal$ to get the eigenvectors $\vec{\zeta}_{-}$ and the eigenvalue 
%
$\exp\left({-i\mu }\right)$. The results are:
%
%
\begin{equation} \exp\left({-i\mu }\right)={1-i\sqrt {3} \over 2}~~~\Rightarrow \mu =
{\pi  \over 3}, \end{equation}
%
%
\begin{equation}\ \ \ {\vec{\zeta }}_{-}=
{\sqrt {2} \over {3}^{1/4}}\left({\begin{array}{c}1\\
{1+i\sqrt {3} \over 2}\end{array}}\right), \end{equation}
%
%
%
\begin{equation} {x}^{N}=
{\sqrt {2} \over {3}^{1/4}}\left({x+{1 \over 2}{p}_{x}}\right)~~~{\rm a}{\rm n}{\rm d} ~~~{p}^{N}=
{{3}^{1/4} \over \sqrt {2}}{p}_{x}, \end{equation}
%
%
\begin{equation} {A}^{-1}=
\left({\begin{array}{cc}{\sqrt {2} \over {3}^{1/4}}&{1 \over \sqrt {2}\ {3}^{1/4}}\\
0&{{3}^{1/4} \over \sqrt {2}}\end{array}}\right), \label{eq:aie}\end{equation}

%
\begin{equation} R=
{A}^{-1}MA=
\left({\begin{array}{cc}{\sqrt {2} \over {3}^{1/4}}&{1 \over \sqrt {2}\ {3}^{1/4}}\\
0&{{3}^{1/4} \over \sqrt {2}}\end{array}}\right)\left({\begin{array}{cc}1&1\vphantom{{{3}^{1/4} \over \sqrt {2}}}\\
-1&0\vphantom{{{3}^{1/4} \over \sqrt {2}}}\end{array}}\right)\left({\begin{array}{cc}{{3}^{1/4} \over \sqrt {2}}&-{1 \over \sqrt {2}\ {3}^{1/4}}\\
0&{\sqrt {2} \over {3}^{1/4}}\end{array}}\right)=
\left({\begin{array}{cc}{1 \over 2}&{\sqrt {3} \over 2}\\
-{\sqrt {3} \over 2}&{1 \over 2}\end{array}}\right). \label{eq:aiem}\end{equation}
We can compute the invariant using $A^{-1}$, it is the so-called Courant-Snyder invariant: 
%
\begin{equation} {\varepsilon =
\left({{\sqrt {2} \over {3}^{1/4}}x+{1 \over \sqrt {2}\ {3}^{1/4}}{p}_{x}}\right)}^{2}+{\sqrt {3} \over 2}{p}_{x}^{2}=
{2 \over \sqrt {3}}{x}^{2}+2{1 \over \sqrt {3}}x{p}_{x}+{2 \over \sqrt {3}}{p}_{x}^{2} ,\end{equation}
from which we see that
%
\begin{equation} \gamma =
\beta =
{2 \over \sqrt {3}}~~~~~{\rm a}{\rm n}{\rm d} ~~~~~\alpha =
{1 \over \sqrt {3}}. \end{equation}
We notice that our method has produced in \Eref{eq:aie}, accidentally, the canonical transformation of \Eref{eq:csca}. We can verify \Eref{eq:aib} using the lattice function extracted from the invariant:%
%
\begin{equation} {A}_{C-S}^{-1}{A}_{{\rm c}{\rm o}{\rm m}{\rm p}{\rm u}{\rm t}{\rm e}{\rm d} }=
\left({\begin{array}{cc}{\left({{\sqrt {3} \over 2}}\right)}^{1/2}&0\\
{\left({{1 \over 2\sqrt {3}}}\right)}^{1/2}&{\left({{2 \over \sqrt {3}}}\right)}^{1/2}\end{array}}\right)\left({\begin{array}{cc}{\left({{\sqrt {3} \over 2}}\right)}^{1/2}&{-\left({{1 \over 2\sqrt {3}}}\right)}^{1/2}\\
0&{\left({{2 \over \sqrt {3}}}\right)}^{1/2}\end{array}}\right)=
\left({\begin{array}{cc}{\sqrt {3} \over 2}&-{1 \over 2}\\
{1 \over 2}&{\sqrt {3} \over 2}\end{array}}\right).\label{eq:aibe}\end{equation}
The reader can compare the rotations of \Eref{eq:aibe} and  \Eref{eq:aib} using $\alpha =  1/\sqrt{3}$.

\subsection{The actual rotation and phasors from the code}\label{sec:rot}

According to \Eref{eq:acs}, i.e.,
%
\[ M=
A\circ R\circ {A}^{-1} \]
the normal from is a rotation. For example, we run the \vn{z_guido_lecture.f90}, we get:

\begin{verbatim} 
 Normal form: Rotation

           2  Dimensional map

          1, NO =    2, NV =    2, INA =  206
 *********************************************

    I  COEFFICIENT          ORDER   EXPONENTS
      NO =     2      NV =     2
   1  0.3685615444773485       0.000000000000000       1  0
   1  0.9296033497855252       0.000000000000000       0  1
    -2   0.000000000000000       0.000000000000000       0  0

          1, NO =    2, NV =    2, INA =  207
 *********************************************

    I  COEFFICIENT          ORDER   EXPONENTS
      NO =     2      NV =     2
   1 -0.9296033497855257       0.000000000000000       1  0
   1  0.3685615444773486       0.000000000000000       0  1
    -2   0.000000000000000       0.000000000000000       0  0
\end{verbatim} 

This rotation can be  diagonalized into ``the phasors'' basis:
%
\begin{equation} \Lambda (\mu )=
{C}^{-1}\circ R\circ C~~~~{\rm w}{\rm h}{\rm e}{\rm r}{\rm e}~~~~~\Lambda =
D(-\mu )=
\left({\begin{array}{cc}\exp\left({-i\mu }\right)&0\\
0&\exp\left({i\mu }\right)\end{array}}\right).\label{eq:ster} \end{equation}

$D(-\mu )$ is the matrix in Dr.~Sterbini's lecture. The FPP package has a different definition for the~phasors: the first and second phasors are inverted leading to this unfortunate alternate convention.  The matrix $C^{-1}$ of the FPP package is given by:
%
\begin{equation} {C}^{-1}=
{1 \over \sqrt {2}}\left({\begin{array}{cc}1&i\\
1&-i\end{array}}\right).\label{eq:ci}\end{equation}
It should be pointed that this is not the usual definition of FPP as indicated by the  line \vn{n_cai=-i_} below. Indeed if the assignment  \vn{n_cai=-i_} is omitted then there is no $\sqrt {2}$ in \Eref{eq:ci}---FPP's default choice.
The result from the code is:

\begin{verbatim} 
n_cai=-i_     !  this puts the sqrt(2) in the phasor 
                  ! rather than phasor =x +ip (My usual choice)
                .
                .
                .
 D(-mu) (because of FPP different phasors' definition)

          2  Dimensional map

         1, NO =    2, NV =    2, INA =  221
*********************************************

   I  COEFFICIENT          ORDER   EXPONENTS
     NO =     2      NV =     2
  1  0.3685615444773486     -0.9296033497855254       1  0
   -1   0.000000000000000       0.000000000000000       0  0

         1, NO =    2, NV =    2, INA =  222
*********************************************

   I  COEFFICIENT          ORDER   EXPONENTS
     NO =     2      NV =     2
  1  0.3685615444773486      0.9296033497855254       0  1
   -1   0.000000000000000       0.000000000000000       0  0


\end{verbatim} 

\subsection{The Courant-Snyder loop}\label{sec:csloop}

In general we need to examine an accelerator at several locations. For example, the equations of motion assume a knowledge of physics everywhere in space. In practice, in a tracking code, propagation is known at a finite number of surface of sections around the machine, namely at every integration step. In an experimental setting, knowledge of  phase space is restricted by the finite number of beam position monitors or BPM.  The Courant-Snyder loop shows how one can express the motion from point  1 to point~2 in terms of the normalised map. This is shown in \Fref{fig:cs}.

\begin{figure}
\centering \includegraphics[width=.5\linewidth]{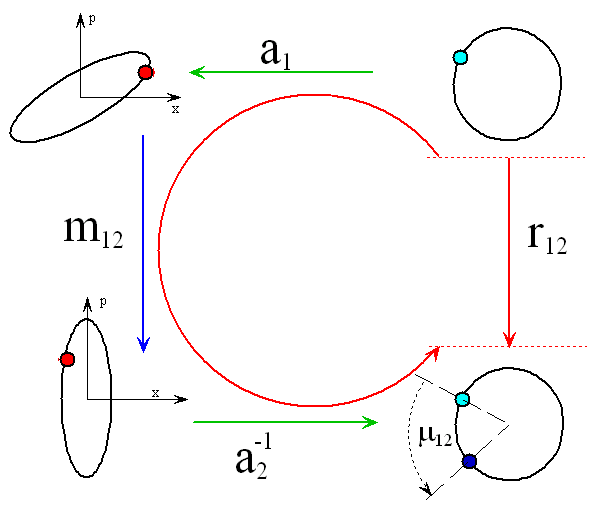}
\caption{The Courant-Snyder loop}
\label{fig:cs}
\end{figure}

One can put   \Fref{fig:cs}  in algebraic form: 
%
\begin{equation} {r}_{12}=
{a}_{2}^{-1}\circ {m}_{12}\circ {a}_{1}.\label{eq:m12}\end{equation}
The maps $m_1$ and $m_2$ are normalised by $a_1$ and $a_2$. In other words, we have
%
\begin{equation} r=
{a}_{1}^{-1}\circ {m}_{1}\circ {a}_{1}~~~{\rm a}{\rm n}{\rm d} ~~~r=
{a}_{2}^{-1}\circ {m}_{2}\circ {a}_{2}.\label{eq:a1a2}\end{equation}
In \Eref{eq:a1a2},  it is assumed that the one-turn map at position $i=1,2$ is normalised by a given method, for example   the linear  Courant-Snyder method of \Eref{eq:csc}. The only important thing is that the map, linear or nonlinear,  is normalised by a well-defined method that depends on the map alone. Then \Eref{eq:m12} defines the phase advance between positions 1 and 2: it is the angle of the rotation $r_{12}$.

The following code fragment implements \Fref{fig:cs} or equivalently \Eref{eq:a1a2}.

\begin{verbatim}

f=>cell%start
do i=1,cell%n
 call propagate(cell,xs,+state,fibre1=i,fibre2=i+1)

 a=xs  ! creates tracked canonical transformation
 xs0=xs ! Saves closed orbit at position i
call c_canonise(a,a_cs,phase=phase) ! A = A_cs o R(phase)
 
 
xs=xs0+a_cs ! updates the  real_8 ``xs''  with the ``canonised transformation'' a_cs

f=>f%next
enddo

\end{verbatim}

\subsubsection{Analytical Courant-Snyder phase advance}\label{sec:anaph}
The above code fragment from z_guido_lecture.f90 displays a bare bone Courant-Snyder loop. Just for fun, we can compute the phase advance assuming the usual simple Hamiltonian for the code:
 %
\begin{equation} H=
{{p}_{x}^{2} \over 2}+k(s){{x}^{2} \over 2}\label{eq:hsa}.\end{equation}
We compute the matrix of the  map $m_{12}$, i.e. $M_{12}$, assuming that position 1 corresponds to position $s$ and that position 2 is at $s+ds$. Then the matrix for this map is to order $ds$:
%
\begin{equation} {M}_{12}=
\left({\begin{array}{cc}1&ds\\
-kds&1\end{array}}\right).\label{eq:mat12} \end{equation}
Then using \Eref{eq:mat12}, i.e., 
%
\begin{equation} {M}_{12}=
{A}_{2}{R}_{12}{A}_{1}^{-1}\Rightarrow {A}_{2}=
{M}_{12}{A}_{1}{R}_{12}^{-1},\label{eq:M12}\end{equation}
we can write:
%
\[ {M}_{12}=
\left({\begin{array}{cc}1&ds\\
-kds&1\end{array}}\right)\ \ \ \ \ {A}_{1}=
\left({\begin{array}{cc}\sqrt {{\beta }_{s}}&0\\
-{\alpha }_{s}/\sqrt {{\beta }_{s}}&1/\sqrt {{\beta }_{s}}\end{array}}\right)\ \ \ \ \ {R}_{12}^{-1}=
\left({\begin{array}{cc}1&-d\mu \\
d\mu &1\end{array}}\right)\ \ \ \ \ \ ~\ \]
\[ \Downarrow \]
%
$${A}_{2}=
\left({\begin{array}{cc}\vphantom{\mbox{\LARGE ${\sqrt {{\beta }_{s}} \over \sqrt {{\beta }_{s}}}$}}\sqrt {{\beta }_{s}}-ds\ {\alpha }_{s}/\sqrt {{\beta }_{s}}&-d\mu \sqrt {{\beta }_{s}}+ds/\sqrt {{\beta }_{s}}\\
\vphantom{\mbox{\LARGE ${\sqrt {{\beta }_{s}} \over \sqrt {{\beta }_{s}}}$}}-{\alpha }_{s}/\sqrt {{\beta }_{s}}+ds/{\beta }_{s}^{3/2}-k\sqrt {{\beta }_{s}}ds&1/\sqrt {{\beta }_{s}}+ds\ {\alpha }_{s}/{\beta }_{s}^{3/2}\end{array}}\right)\ +\ O{(ds,d\mu )}^{2}$$
The Courant-Snyder definition for $A$ implies that $A_{12}=0$. From this condition we deduce that:
%
\begin{equation} -d\mu \sqrt {{\beta }_{s}}+ds/\sqrt {{\beta }_{s}}=
0\ \Longrightarrow {d\mu  \over ds}=
{1 \over {\beta }_{s}}.\label{eq:phdef} \end{equation}
\Eref{eq:phdef}  is the famous formula for the phase advance in 1-d-f in an accelerator.

\subsubsection{Analytical evolution of lattice functions }\label{sec:analat}

Using the infinitesimal version of the map $m_{12}$ of  \Eref{eq:mat12}, we can also derive the $s$ dependence of the~lattice functions.  We start with 
\[ {\varepsilon }_{1}=
{\varepsilon }_{1}\circ {m}_{1}=
>{\varepsilon }_{1}=
{\varepsilon }_{1}\circ \underbrace{{m}_{12}^{-1}\circ {m}_{2}\circ {m}_{12}}\limits_{{m}_{1}}^{},\]
from which we get
%
\begin{equation}  {\varepsilon }_{2}=
{\varepsilon }_{1}\circ {m}_{12}^{-1}.\label{eq:epsm12} \end{equation}
Using again \Eref{eq:mat12} for $m_{12}$, we compute  \Eref{eq:epsm12}  to first order in $ds$:
%
\begin{equation} {d\beta  \over ds}=
-2\alpha \ \ \ \ \ {d\gamma  \over ds}=
2\alpha k\ \ \ \ \ {d\alpha  \over ds}=
\beta k-\gamma .\ \ \ \label{eq:dlatds} \end{equation}

\subsubsection{Some result from the tracking loop}\label{sec:res}

Using the formulae of \Eref{eq:dlatds}, we  derive the value of the beta-function at the end of a drift of length $L$ in terms of the initial lattice functions:
%
\begin{equation} {\beta }_{L}=
{\gamma }_{0}{L}^{2}-2{\alpha }_{0}L+{\beta }_{0}.\label{eq:betad}\end{equation}
We can test \Eref{eq:betad} by comparing the with the result of \Eref{eq:epsm12}  when $m_{12}$ is a full drift. Here is a result from the file \vn{twiss_from_guido.txt}:

\begin{verbatim}
  end of Magnet BEND                    
 Phase advance 
          1, NO =    2, NV =    2, INA =  412
 *********************************************

    I  COEFFICIENT          ORDER   EXPONENTS
      NO =     2      NV =     2
   0  0.6731339404229452       0.000000000000000       0  0
    -1   0.000000000000000       0.000000000000000       0  0
 Invariant 

          1, NO =    2, NV =    2, INA =  411
 *********************************************
    
    I  COEFFICIENT          ORDER   EXPONENTS
      NO =     2      NV =     2
   2   1.592892522601233       0.000000000000000       2  0
   2  -1.803500025004616       0.000000000000000       1  1
   2   1.138277102391687       0.000000000000000       0  2
    -3   0.000000000000000       0.000000000000000       0  0
    
    
    
    
  Beta at next element computed    1.51178141412459    <-------------  beta from analytical
   Phase at next element   0.694900568502571  <----------------|                        |                                         |    
  end of Magnet L15                                            |                        |
 Phase advance                                                 |                        |
                                                               |                        |
          1, NO =    2, NV =    2, INA =  412                  |                        |
 *********************************************                 |                        |
                                                               |                        |
    I  COEFFICIENT          ORDER   EXPONENTS                  |                        |
      NO =     2      NV =     2                               |                        |
   0  0.6949005685025705       0.000000000000000       0  0 <--|                        ^
    -1   0.000000000000000       0.000000000000000       0  0                           ^
 Invariant                                                                              |
                                                                                        |
          1, NO =    2, NV =    2, INA =  411                                           |
 *********************************************                                          |
                                                                                        |
    I  COEFFICIENT          ORDER   EXPONENTS                                           | 
      NO =     2      NV =     2                                                        |
   2   1.592892522601233       0.000000000000000       2  0                             |
   2  -2.373272264504499       0.000000000000000       1  1                             |
   2   1.511781414124592       0.000000000000000       0  2    <--------  beta from TPSA
    -3   0.000000000000000       0.000000000000000       0  0
  end of Magnet L16                    
\end{verbatim}

Using \Eref{eq:betad}, one can also compute the phase advance analytically through a drift:
%
\begin{equation} \nu (L)=
{\nu }_{0}+\int_{0}^{L}{ds \over {\gamma }_{0}{s}^{2}-2{\alpha }_{0}s+{\beta }_{0}}=
{\tan}^{-1}\left({{\gamma }_{0}L-{\alpha }_{0}}\right)+{\tan}^{-1}\left({{\alpha }_{0}}\right).\label{eq:dmua}\end{equation}

The analytical results of this section assume mid-plane symmetry and the Hamiltonian of \Eref{eq:hsa}.

\section{Nonlinearities  }\label{sec:nonlin}

The beauty of a map based theory is three-fold:
\begin{enumerate}
\item Thanks to TPSA, it can be implemented in a tracking code (integrator type), with little effort.
\item The linear theory can be extended to the nonlinear regime through the concept of the map ${\cal M}=\circ M$ which acts on functions.
\item It can be extended to spin tracking because the normal form for spin is also a planar rotation.
\end{enumerate}

In this final section, we do a little nonlinear analytical calculation followed with a numerical computation using the library FPP. FPP can easily carry the calculation to higher order. This is important the case of sextupoles: tune shifts with amplitude only appear at the second order in strength.

\subsection{Phasors: from linear to nonlinear }\label{sec:phasors}

We have already seen the following equation for the invariant:
%
\begin{equation} \varepsilon =
{\cal M}\varepsilon =
\varepsilon \circ M.\ \ \ \ \label{eq:invc2}\end{equation}

The first step to solve \Eref{eq:invc2} is to examine the nonlinear eigenfunctions of rotation, the so-called phasors. The procedure to normalise a nonlinear map will boil down to re-expressing the map in terms of phasors through a Hamiltonian representation of the map. The phasor eigenfunctions are the same as the objects represented by the component vector of \Eref{eq:phasv}. 

\subsubsection{Phasors defined as the eigenfunctions of $\circ R$ }\label{sec:phaseig}

Using \Eref{eq:aiem}, we can guess the form of the map ${\cal R} = \circ R$ in terms of the Poisson bracket
%
\begin{equation} {\cal R}=
\ \exp\left({-{\mu  \over 2}\left[{{x}^{2}+{p}_{x}^{2},\ \ \ }\right]}\right)=
\exp\left({-{\mu  \over 2}:{x}^{2}+{p}_{x}^{2}:}\right) \label{eq:rlie}  \end{equation}
%
%
\[ {\rm w}{\rm h}{\rm e}{\rm r}{\rm e}~~~:h:f=
\left[{h,f}\right]=
{\partial h \over \partial x}{\partial f \over \partial {p}_{x}}-{\partial h \over \partial {p}_{x}}{\partial f \over \partial x}. \]
We can let the operator $:{x}^{2}+{p}_{x}^{2}:$ act of the function $x+i p_x$:
%
\begin{equation} -{\mu  \over 2}:{x}^{2}+{p}_{x}^{2}:\left({x+i{p}_{x}}\right)=
-i\mu \left({x+i{p}_{x}}\right)\ \Rightarrow \ {e}^{-{\mu  \over 2}:{x}^{2}+{p}_{x}^{2}:}\left({x+i{p}_{x}}\right)=
{e}^{-i\mu }\left({x+i{p}_{x}}\right).\label{eq:eigmuimp}\end{equation}
Then writing  $x$ as
%
\[ x=
{1 \over 2}\left\{{\left({x+i{p}_{x}}\right)\ +\ \left({x-i{p}_{x}}\right)}\right\}, \]
we get
%
%
\begin{equation} \exp\left({-{\mu  \over 2}:{x}^{2}+{p}_{x}^{2}:}\right)x=
\cos\left({\mu }\right)x+\sin\left({\mu }\right){p}_{x}.\label{eq:rx}\end{equation}
Then writing $p_x$ as 
%
\[ {p}_{x}=
{1 \over 2i}\left\{{\left({x+i{p}_{x}}\right)\ -\ \left({x-i{p}_{x}}\right)}\right\}, \]
we get 
%
\begin{equation} \exp\left({-{\mu  \over 2}:{x}^{2}+{p}_{x}^{2}:}\right){p}_{x}=
\cos\left({\mu }\right){p}_{x}-\sin\left({\mu }\right)x.\label{eq:rliepx}\end{equation}
Thus we conclude that the map $\cal R$ is associated to the matrix $R$. However $\cal R$ can act on nonlinear functions. For example, if we expand a function in terms of %
%
${h}_{\mp}=
{1 \over \sqrt {2}}\left({x\pm i{p}_{x}}\right)$,
%
%
\[ f=
\sum\limits_{{m}_{1},{m}_{2}}^{} {F}_{\vec{m}}{h}_{+}^{{m}_{1}}{h}_{-}^{{m}_{2}}, \]
and let $\cal R$ act on it, the answer is very simple:
%
\begin{equation} {\cal R}f=
f\circ R=
\sum\limits_{{m}_{1},{m}_{2}}^{} {F}_{\vec{m}}{\left({{h}_{+}\circ R}\right)}^{{m}_{1}}{\left({{h}_{-}\circ R}\right)}^{{m}_{2}}=
\sum\limits_{{m}_{1},{m}_{2}}^{} {e}^{i\mu \left({{m}_{2}-{m}_{1}}\right)}{F}_{\vec{m}}{h}_{+}^{{m}_{1}}{h}_{-}^{{m}_{2}}.\label{eq:fcircr} \end{equation}
The eigenfunctions ${h}_{+}$ and ${h}_{-}$ can be used to define the so-called action-angle variables $\phi$ and $J$:
%
\begin{equation}~{\rm i}{\rm f}~J=
{h}_{+}{h}_{-}~~~~~x=
\sqrt {2J}\cos\left({\phi }\right)~~~{\rm a}{\rm n}{\rm d} ~~~{p}_{x}=
-\sqrt {2J}\sin\left({\phi }\right).\ \label{eq:xpj} \end{equation}
 $\phi$ and $J$ form a canonical pair: the reader can check that $[x,p_x]=1$  implies  $[\phi,J]=1$.

\subsubsection{Expressing the maps with phasors }\label{sec:phasorsmap }

We want to use the monomials in \Eref{eq:fcircr} to re-express any non-linear map. We first compute the Poisson bracket between $h_{+}$ and $h_{-}$:
%
\begin{equation} \left[{{h}_{+},{h}_{-}}\right]=
\left[{{1 \over \sqrt {2}}\left({x+i{p}_{x}}\right),{1 \over \sqrt {2}}\left({x-i{p}_{x}}\right)}\right]=
-i.\label{eq:pbp} \end{equation}
We can now define a new operator which assumes that the functions are in phasors:
%
\begin{equation} {}_{\star}^{\star}f{}_{\star}^{\star}g\ =
\ -i\left[{f,g}\right]. \label{eq:pbphasor} \end{equation}
Then the map $\cal R$ of \Eref{eq:rlie} can be written as:
%
\begin{equation} {\cal R}=
\ \exp\left({-{\mu  \over 2}\left[{{x}^{2}+{p}_{x}^{2},\ \ \ }\right]}\right)=
\exp\left({-\mu {}_{\star}^{\star}{h}_{+}{h}_{-}{}_{\star}^{\star}}\right).\label{eq:rliep}  \end{equation}
To show how this can work in general, we can look at a map which is a rotation followed by an octupole kick. First the rotation
%
\[ {\vec{z}}^{i}=
R{\vec{z}}^{0},\]
and the octupole kick
%
\begin{equation} {\rm M}{\rm a}{\rm p}~O:~~~~~~~{x}^{1}=
{x}^{i}~~~~{\rm a}{\rm n}{\rm d} ~~~~{p}_{x}^{1}=
{p}_{x}^{i}-{k}_{o}{{x}^{i}}^{3}.\label{eq:mapexa} \end{equation}
The reader can check that the map acting on functions, ${\cal O} =\circ O$, is given by:
%
\begin{equation} {\cal O}=
\exp\left({:-{{k}_{o} \over 4}{x}^{4}:}\right).\label{eq:octmap}\end{equation}
Calling the total map $M$, we have 
%
\begin{equation} M=
O\circ R.\label{eq:rotoct} \end{equation}
Let us act on an arbitrary function $f$,
%
\[ {\cal M}f=
f\circ O\circ R=
{\cal O}f\circ R=
{\cal R}{\cal O}f, \]
from which we notice that the order of concatenation is opposite to that of phase space maps. This is not surprising since we saw in the linear case that the transposed matrix acts on functions (see \Eref{eq:mct}).

The total map can be written as
%
%
\begin{equation} {\cal M}={\cal R}{\cal O}=
{\cal R}\exp\left({:-{{k}_{o} \over 4}{x}^{4}:}\right).\label{eq:mro}\end{equation}
Since the linear part is already a rotation, we  express the nonlinear part ---the octupole in this case--- in terms of the phasors of \Eref{eq:pbp} which are eigenfunctions of $\cal R$:
%
\begin{equation} {\cal M}={\cal R}{\cal O}=
\exp\left({-\mu {}_{\star}^{\star}{h}_{+}{h}_{-}{}_{\star}^{\star}}\right)\exp\left({{}_{\star}^{\star}-{{k}_{o} \over 16}\left({{h}_{+}^{4}+{h}_{-}^{4}+4{h}_{+}^{3}{h}_{-}+4{h}_{+}{h}_{-}^{3}+6{h}_{+}^{2}{h}_{-}^{2}}\right){}_{\star}^{\star}}\right).\label{eq:mroe}\end{equation}

\subsection{Nonlinear normal form: octupole example and its tune }\label{sec:nonlinearnf}

Usually we start with a map that has already been put into a linear normal form, i.e., its linear part is a~rotation. Therefore, the octupole example of \Eref{eq:mro} is a good example. Let us assume that the~following can be done:
%
\[ {{\cal R}}_{1}=
{{\cal A}}_{1}{\cal M}{{\cal A}}_{1}^{-1}=
\exp\left({-\mu {}_{\star}^{\star}{h}_{+}{h}_{-}{}_{\star}^{\star}}\right)\exp\left({{}_{\star}^{\star}-{3{k}_{o} \over 8}{h}_{+}^{2}{h}_{-}^{2}{}_{\star}^{\star}}\right) \]
%
\begin{equation} \hphantom{{{\cal R}}_{1}=
{{\cal A}}_{1}{\cal M}{{\cal A}}_{1}^{-1}}=
\exp\left({{}_{\star}^{\star}-\mu {h}_{+}{h}_{-}-{3{k}_{o} \over 8}{h}_{+}^{2}{h}_{-}^{2}{}_{\star}^{\star}}\right).\label{eq:mro1}\end{equation}
The reader will check that the exponents in \Eref{eq:mro1} commute and therefore can be added.

In \Eref{eq:mro1}, we assumed that the terms with unequal power in phasors are removed by canonical transformation. We are left with terms with equal powers in phasors, i.e., a rotation. We can let the map ${\cal R}_{1}$ on the phasors $h_{\mp}$:
%
\begin{equation} {{\cal R}}_{1}{h}_{\mp}=
\exp\left({{}_{\star}^{\star}-\mu {h}_{+}{h}_{-}-{3{k}_{o} \over 8}{h}_{+}^{2}{h}_{-}^{2}{}_{\star}^{\star}}\right){h}_{\mp}=
\exp\left({\pm  i\left\{{\mu +{3{k}_{o} \over 4}J}\right\}}\right){h}_{\mp}~~~~~{\rm w}{\rm h}{\rm e}{\rm r}{\rm e}~~J=
{h}_{+}{h}_{-}\label{eq:mro1h}\end{equation}

\subsection{Computation of \protect${{\cal A}}_{1}$ }\label{sec:anonlinearnf}

The map ${{\cal A}}_{1}$ removes the resonant phasors in \Eref{eq:mroe} to leading order in octupole strength $k_o$. We can write  ${{\cal A}}_{1}$ in terms of an operator proportional to $k_o$:
%
\begin{equation} {{\cal A}}_{1}=
\exp\left({\ {}_{\star}^{\star}{k}_{o}F{}_{\star}^{\star}}\right).\label{eq:a1lie}\end{equation}
We now compute its action on the map $\cal M$ of \Eref{eq:mroe} to first order in $k_o$:
%
\[ {{\cal A}}_{1}{\cal M}{{\cal A}}_{1}^{-1}=
{e}^{{}_{\star}^{\star}{k}_{o}F{}_{\star}^{\star}}{\cal R}\exp\left({{}_{\star}^{\star}-{{k}_{o} \over 16}\left({{h}_{+}^{4}+{h}_{-}^{4}+4{h}_{+}^{3}{h}_{-}+4{h}_{+}{h}_{-}^{3}+6{h}_{+}^{2}{h}_{-}^{2}}\right){}_{\star}^{\star}}\right){e}^{-{}_{\star}^{\star}{k}_{o}F{}_{\star}^{\star}}, \]
%
%
\[ \hphantom{{{\cal A}}_{1}{\cal M}{{\cal A}}_{1}^{-1}}=
{\cal R}{e}^{{{\cal R}}^{-1}{}_{\star}^{\star}{k}_{o}F{}_{\star}^{\star}{\cal R}}\exp\left({{}_{\star}^{\star}-{{k}_{o} \over 16}\left({{h}_{+}^{4}+{h}_{-}^{4}+4{h}_{+}^{3}{h}_{-}+4{h}_{+}{h}_{-}^{3}+6{h}_{+}^{2}{h}_{-}^{2}}\right){}_{\star}^{\star}}\right){e}^{-{}_{\star}^{\star}{k}_{o}F{}_{\star}^{\star}}, \]
using %
%
\[ {{\cal R}}^{-1}{}_{\star}^{\star}{k}_{o}F{}_{\star}^{\star}{\cal R} = {}_{\star}^{\star}{k}_{o}{{\cal R}}^{-1}F{}_{\star}^{\star}= {}_{\star}^{\star}{k}_{o}F\circ {{\cal R}}^{-1}{}_{\star}^{\star} \]
%
%
\begin{equation} \hphantom{{{\cal A}}_{1}{\cal M}{{\cal A}}_{1}^{-1}}\approx {\cal R}\exp\left({{}_{\star}^{\star}{k}_{o}\left({F\circ {R}^{-1}-F}\right)-{{k}_{o} \over 16}\left({{h}_{+}^{4}+{h}_{-}^{4}+4{h}_{+}^{3}{h}_{-}+4{h}_{+}{h}_{-}^{3}+6{h}_{+}^{2}{h}_{-}^{2}}\right){}_{\star}^{\star}}\right) + O({k}_{o}^{2}) \label{eq:ota1}\end{equation}
The trick is to choose the function $F$  to remove the resonant terms:
%
\[ F=
-{\left({1-{{\cal R}}^{-1}}\right)}^{-1}\left({{h}_{+}^{4}+{h}_{-}^{4}+4{h}_{+}^{3}{h}_{-}+4{h}_{+}{h}_{-}^{3}}\right) \]
%
%
\begin{equation} \hphantom{F}=
-{1 \over 16}\left\{{{{h}_{+}^{4} \over 1-{e}^{i4\mu }}+{{h}_{-}^{4} \over 1-{e}^{-i4\mu }}+{4{h}_{+}^{3}{h}_{-} \over 1-{e}^{i2\mu }}+{4{h}_{+}{h}_{-}^{3} \over 1-{e}^{-i2\mu }}}\right\}.\label{eq:solf} \end{equation}
In \Eref{eq:solf} we used  the fact that all monomials in phasors are eigenfunctions of  ${\cal R}^{-1}$.

\subsection{Computation of the  Invariant}\label{sec:anonlinearinv}

We start with the map $\cal M$
%
\begin{equation} {\cal M}=
{{\cal A}}_{1}^{-1}{{\cal R}}_{1}{{\cal A}}_{1},\label{eq:maro1}\end{equation}
As in \Eref{eq:csa}, the invariant is gotten by acting with ${\cal A}_1^{-1}$ on the half-radius $J=h_+ h_-$:
%
%
\[ {\varepsilon  \over 2}=
{{\cal A}}_{1}^{-1}{h}_{+}{h}_{-}=
{h}_{+}{h}_{-}-{k}_{o}{}_{\star}^{\star}F{}_{\star}^{\star}{h}_{+}{h}_{-}+O({k}_{o}^{2})\cdots \]
%
%
\begin{equation} \ =
{h}_{+}{h}_{-}-i{{k}_{o} \over 4}\left\{{{{h}_{+}^{4} \over 1-{e}^{i4\mu }}-{{h}_{-}^{4} \over 1-{e}^{-i4\mu }}+{2{h}_{+}^{3}{h}_{-} \over 1-{e}^{i2\mu }}-{2{h}_{+}{h}_{-}^{3} \over 1-{e}^{-i2\mu }}}\right\}\cdots\label{eq:invexp} \end{equation}

Here are the numerical results from the program \vn{z_rotation_octupole.f90}.

{\small 
\begin{verbatim}
 Map in regular Cartesian variables

           2  Dimensional map

 Properties, NO =    4, NV =    2, INA =  157
 *********************************************

   1  0.7154976714602418       0.000000000000000       1  0
   1  0.6986151173106488       0.000000000000000       0  1


 Properties, NO =    4, NV =    2, INA =  156
 *********************************************

   1 -0.6986151173106488       0.000000000000000       1  0
   1  0.7154976714602418       0.000000000000000       0  1
   3 -0.3662896726669607       0.000000000000000       3  0
   3  -1.072940609789786       0.000000000000000       2  1
   3  -1.047623996379843       0.000000000000000       1  2
   3 -0.3409682473807201       0.000000000000000       0  3

 zeroth order tune   0.123100000000000
 
 Lie polynomial of A_1 in phasors

 Properties, NO =    4, NV =    2, INA =  256
 *********************************************

   4 -0.3125000000000000E-01 -0.7462700689329130E-03   4  0
   4 -0.1250000000000000     -0.1280207180125487       3  1
   4 -0.1250000000000000      0.1280207180125487       1  3
   4 -0.3125000000000000E-01  0.7462700689329174E-03   0  4

  Analytical Results
  Coefficient of h_+^4
 (-3.125000000000000E-002,-7.462700689328990E-004)
  Coefficient of h_+^3 h_-
 (-0.125000000000000,-0.128020718012549)
 Nonlinear invariant in phasors

 Properties, NO =    4, NV =    2, INA =  256
 *********************************************

   2   1.000000000000000       0.000000000000000       1  1
   4  0.2985080275731647E-02 -0.1250000000000000       4  0
   4  0.2560414360250975     -0.2500000000000000       3  1
   4  0.2560414360250974      0.2500000000000000       1  3
   4  0.2985080275731647E-02  0.1250000000000000       0  4

  Analytical Results
  Coefficient of h_+^4
 (2.985080275731596E-003,-0.125000000000000)
  Coefficient of h_+^3 h_-
 (0.256041436025097,-0.250000000000000)
 
 Nonlinear part of the Rotation

           2  Dimensional map

 Properties, NO =    4, NV =    2, INA =  184
 *********************************************

   1   1.000000000000000       0.000000000000000       1  0
   3  0.3749999999999998       0.000000000000000       2  1
   3  0.3749999999999999       0.000000000000000       0  3


 Properties, NO =    4, NV =    2, INA =  183
 *********************************************

   1  0.9999999999999999       0.000000000000000       0  1
   3 -0.3749999999999998       0.000000000000000       3  0
   3 -0.3750000000000002       0.000000000000000       1  2
   
 Lie polynomial of the nonlinear part of the rotation

 Properties, NO =    4, NV =    2, INA =  256
 *********************************************

   4 -0.3750000000000000       0.000000000000000       2  2

  Analytical Results
  Coefficient of h_+^ 2h_-^2 =  -0.375000000000000
\end{verbatim}
}

\appendix

\section{Programme z_rotation_octupole.f90}\label{sec:zrotationoctupole}

The programme below was used to check the analytical results.

{\small 
\begin{verbatim}
program Octupole_kick
use pointer_lattice
implicit none
integer no,np 
real(dp) mu
type(c_damap) id,one_turn_map
type(c_taylor) hr
type(c_normal_form) normal_form
type(c_vector_field) fo
type(internal_state) state
real(dp) Q_0 
 call c_print_eps(1.d-5)
!!!  print format is cleaner for TPSA
longprint=.false.
!!! removes unwanted print statement
c_verbose=.false.
lielib_print(4)=0

n_cai=-i_     !  this puts the sqrt(2) in the phasor 
              ! rather than phasor =x +ip (My usual choice)
 call ptc_ini_no_append

state=only_2d0  
no=4     

call init_all(state,no,np)
 
!   create these TPSA objects
call alloc(id,one_turn_map) 
call alloc(normal_form)
call alloc(hr)
call alloc(fo)

mu=twopi*.1231d0


!  Lie operator of rotation
hr=-(mu/2)*((1.d0.mono.'20')+(1.d0.mono.'02'))

fo=getvectorfield(hr)
one_turn_map=exp(fo)

!  Lie operator of octupole 1/4 x^4
hr=-(0.25d0.mono.'40')

fo=getvectorfield(hr)
id=exp(fo)
!! Map of Octupole o Rotation 
 one_turn_map= id*one_turn_map

Write(6,*) "Map in regular Cartesian variables "
call print(one_turn_map)

call c_normal(one_turn_map,normal_form)  
Q_0=normal_form%tune(1)
write(6,*) "zeroth order tune ",q_0

fo=log(normal_form%atot)
hr=getpb(fo)
hr=hr*c_phasor()
 Write(6,*) "Lie polynomial of A_1 in phasors"
call print(hr)
write(6,*) " Analytical Results "
write(6,*) " Coefficient of h_+^4 "
write(6,*) -1.d0/16.d0/(1.d0-exp(i_*4*mu))
write(6,*) " Coefficient of h_+^3 h_-"
write(6,*) -4.d0/16.d0/(1.d0-exp(i_*2*mu))


Write(6,*) "Nonlinear invariant in phasors"
hr=(((1.d0.mono.'20')+(1.d0.mono.'02')))/2.d0
hr=hr*normal_form%atot**(-1)
hr=hr*c_phasor()
call print(hr)

write(6,*) " Analytical Results "
write(6,*) " Coefficient of h_+^4 "
write(6,*) -i_/4.d0/(1.d0-exp(i_*4*mu))
write(6,*) " Coefficient of h_+^3 h_-"
write(6,*) -i_/2.d0/(1.d0-exp(i_*2*mu))

id=normal_form%atot**(-1)*one_turn_map*normal_form%atot

Write(6,*) "Nonlinear part of the Rotation "

id=(id.sub.1)**(-1)*id
call print(id)

fo=log(id)
hr=getpb(fo)
hr=hr*c_phasor()
Write(6,*) "Lie polynomial of the nonlinear part of the rotation"
call print(hr)
write(6,*) " Analytical Results "
write(6,*) " Coefficient of h_+^ 2h_-^2 = ",  -3.d0/8.d0

end program Octupole_kick
\end{verbatim}
}

\section{Program z_guido_lecture.f90 }\label{sec:zguidolecture}

\begin{verbatim}

program guido_twiss
use pointer_lattice
implicit none
type(layout),pointer :: cell
integer no,np,mf,i,k
real(dp) closed_orbit(6),flip_phase,betax,gammax,alphax,dmu
type(probe) xs0
type(probe_8) xs
type(c_damap) id,one_turn_map,a_cs,a0,a1,a2,a,s,rot,D_mu,P,P_Guido,Omega,J
type(c_taylor) phase(3),phase_one_turn_map(3),r2,courant_snyder_invariant
type(c_normal_form) normal_form
type(internal_state) state
type(fibre), pointer :: f,f0
logical :: answer 
type(c_taylor) x_ij(2,2)
character(10) latf(2,2)
real(dp) beta_s,beta_0,Q_0,phase_0,phase_s,dx_co 
 
 call c_print_eps(1.d-5)
!!!  print format is cleaner for TPSA
longprint=.false.
!!! removes unwanted print statement
c_verbose=.false.
lielib_print(4)=0

!!! I choose here Guido's normalisation 
!!! (look at page 28 of his lecture linearopticscalculations.pdf)
!!! however my choice corresponds to 
!!!       (  1/sqrt(2)    i/sqrt(2)  )
!!!   S=  (                          )
!!!       (  1/sqrt(2)   -i/sqrt(2)  )
!!! 
!!!            (  exp(-i mu)        0       )
!!!   D(mu) =  (                            )
!!!            (       0        exp(i mu)   )
!!! 
!!! this is because my first phasor is  
!!!                   phasor =  (x+ip)/sqrt(2) =  sqrt(J)  exp( i phi)  
!!! while Guido  uses phasor = (x-ip)/sqrt(2)  =  sqrt(J)  exp(-i phi)
n_cai=-i_     !  this puts the sqrt(2) in the phasor 
              ! rather than phasor =x +ip (My usual choice)
 
!!!! reading the flat file produced by MAD-X
call ptc_ini_no_append
!call read_lattice_append(M_U,"../../MAD-X/files_for_cas/als_guido/one_cell.txt")
call read_lattice_append(M_U,  &
             "C:\msys64\home\Etienne\MAD-X\files_for_cas\als_guido\one_cell.txt")
cell=>m_u%start



write(6,*) " Do you want delta as a parameter (t or f)? "
write(6,*) " answer false, i.e., f to reproduce Guido's linear results unambiguously "
write(6,*) "   "
write(6,*) " if you write true (t), &
             then delta and a dipole kick will be added as the 3rd and 4th variables "


read(5,*) answer
if(answer) then
! state that produces map with delta dependence but only 1-d-f (x,px,delta)
 state=only_2d0 +delta0   
! Adding one parameter: theta_0 is, which is -L*Bn(1), is set as the fourth parameter. 
! 
 np=1
 f0=>cell%start
 call move_to(cell,f0,"KICK")
 write(6,*) f0%mag%name
 call make_it_knob(f0%magp%bn(1),1,s=-1.d0)
else
 state=only_2d0 ! state that produces map with delta dependence but only 1-d-f (x,px)
 np=0
endif

! finds the closed orbit at position 1 (should be (0,0,0,0,0,0))

closed_orbit(1:6)=0.d0;  
                                                
call find_orbit_x(cell,closed_orbit(1:6),STATE,1.e-8_dp,fibre1=1)  
 
no=2    ! second order so that we can use Invariant  = (x^2 + p^2) o  A^(-1)
 
call init_all(state,no,np)
 
!   create these TPSA objects
call alloc(id,one_turn_map,a_cs,a0,a1,a2,a,s) 
call alloc(rot,D_mu,P,P_Guido,Omega,J)
call alloc(xs)
call alloc(normal_form)
call alloc(r2,courant_snyder_invariant)
call alloc(phase)
call alloc(phase_one_turn_map)

do i=1,2
do k=1,2
 call alloc(x_ij(i,k))
enddo
enddo

!!! Notation of Guido for the Symplectic form as a c_damap
Omega%v(1)=1.d0.cmono.2 ;  Omega%v(2)=-1.d0.cmono.1;


!!!! page 28 of Guido  modulo different phasors
s=ci_phasor()

write(6,*) ;write(6,*) " Guido S modulo flipping the phasors "
call print(s)

xs0=closed_orbit(1:6)   ! xs0 contains orbit and spin 
id=1   !    identity map
xs=id+xs0   !  xs is a probe_8 which can become a Taylor series
  
call propagate(cell,xs,+state,fibre1=1) ! computes one turn map around closed orbit
 
one_turn_map=xs

call kanalnummer(mf,"output.dat")
write(mf,*) " The Ray of type Probe_8"
call print(xs,mf)
write(mf,*) " The Taylor Map of type C_DAMAP"
call print(one_turn_map,mf)
close(mf)

call c_normal(one_turn_map,normal_form,phase=phase_one_turn_map)  
 Q_0=normal_form%tune(1)

 
!  Puts normal_form%atot into Courant-Snyder form even including delta dependence
!  slow routine not use in linear calculation
call c_canonise(normal_form%atot,a_cs)   
 
!!!!! let us check the normal form
! one_turn_map= normal_form%atot o  rotation o  normal_form%atot^-1
rot  = a_cs**(-1) * one_turn_map * a_cs
 
write(6,*) ;write(6,*) "Normal form: Rotation "
 call print(rot)

! D(-mu)  =  S o rot o S^(-1)
write(6,*) ;write(6,*) " D(-mu) (because of FPP different phasors' definition)"
D_mu = S * rot * S**(-1)
 call print(D_mu)
 
! Total complex transformation NONLINEAR P of page 25 and page 31

write(6,*); write(6,*)" Nonlinear P of page 25 and page 31 modulo flipping the phasors"
P =   a_cs * S**(-1)
call print(P);write(6,*);
write(6,*)"Linear part P_Guido of page 25 and page 31 modulo flipping the phasors "
P_Guido = P.cut.(-2)
call print(P_Guido) ! Retain only the linear part
write(6,*); 
! de Moivre formula of page 29 
! (Does not extend to nonlinear but extends to coupling and nonsymplectic)
!!  Please look at my lecture courant_snyder.pdf
write(6,*)"de Moivre formula of page 29 (extends to coupling but not to nonlinearities)"
write(6,*); 

J = a_cs * Omega * a_cs**(-1)
J=J.cut.2  ! Keep linear

write(6,*)"!!!       (  alpha    beta  )"
write(6,*)"!!!   J=  (                 )"
write(6,*)"!!!       (  gamma   -alpha )"

call print(J)

!
! create  radius**2 =  x^2 + p^2
!
r2 = (1.d0.cmono.1)**2 + (1.d0.cmono.2)**2

 
write(6,*)" courant_snyder_invariant = gamma*x^2 + 2*alpha*x*p + beta*p^2 "
courant_snyder_invariant = r2 * a_cs**(-1)

write(6,*)" Extends to nonlinear systems "
call print(courant_snyder_invariant)

!!!  Tracking of lattice functions
 
!!! On page 36, equation 7, Guido does a mathematical manipulation
!!! which absorbs the matrix Omega into the de Moivre matrix J
!!! what he gets are the averages 
!!! <x_i x_j> = <J Omega^-1)_ij courant_snyder_invariant
!!! Please look at my lecture courant_snyder.pdf
do i=1,2
  do k=1,2
    x_ij(i,k)=(1.d0.cmono.i)*(1.d0.cmono.k)
  enddo
enddo 
latf(1,1)="Beta "
latf(2,2)="Gamma "
latf(1,2)="-Alpha "  
Write(6,'(a)') " Guido's choice for the phasors makes x_1 x_2 = J (J=canonical action) "
Write(6,'(a)') " This can be generalised to nonlinear system "; Write(6,'(a)')


do i=1,2
  do k=i,2
     CALL AVERAGE(x_ij(i,k),a_cs,x_ij(i,k))  
    write(6,*) latf(i,k)
    call print(x_ij(i,k))
  enddo
enddo 

write(6,*)  
write(6,*)  " Tracking of lattice functions "
write(6,*)  " Guido's page 48 is my famous Twiss loop ";write(6,*);
write(6,*)  " This works for nonlinear as well as spin "

write(6,*) ' Do you want Courant-Snyder ? If so write "t" ';write(6,*)
write(6,*)"!!!       (  sqrt(beta)                0      )"
write(6,*)"!!!   A = (                                   )"
write(6,*)"!!!       (  -alpha/sqrt(beta)   1/sqrt(beta) )";write(6,*)
write(6,*)" Or "; write(6,*);
write(6,*) ' Do you want Anti-Courant-Snyder ? If so write "f" ';write(6,*)
write(6,*)"!!!       (  1/sqrt(gamma)    -alpha/sqrt(gamma) )"
write(6,*)"!!!   A = (                                      )"
write(6,*)"!!!       (      0                sqrt(gamma)    )";write(6,*)
write(6,*) " write your answer t or f"
read(5,*) answer
!!!!   now the tracking of the Twiss loop starts
if(answer) then
 courant_snyder_teng_edwards=.true.
else
 courant_snyder_teng_edwards=.false.
endif

call c_canonise(a_cs,a_cs)

call kanalnummer(mf,"twiss_from_guido.txt")

write(mf,*) " Initial canonical transformation "
if(answer) then 
 write(mf,*) " Should have A_12=0 "
else
 write(mf,*) " Should have A_21=0 "
if(np>0) then
  write(6,'(a)') " ^^^^^^^^^^^^^^^^^^^^^^^^^^^^^^^^^^^^^^^^^^^^^^^^^^^^^^^^^^^^^^^^^^^"
  write(6,'(a)') " ^ Guido's formula will be wrong with this choice of Phase Advance ^"     
  write(6,'(a)') " ^^^^^^^^^^^^^^^^^^^^^^^^^^^^^^^^^^^^^^^^^^^^^^^^^^^^^^^^^^^^^^^^^^^"
  write(mf,'(a)') " ^^^^^^^^^^^^^^^^^^^^^^^^^^^^^^^^^^^^^^^^^^^^^^^^^^^^^^^^^^^^^^^^^^^"
  write(mf,'(a)') " ^ Guido's formula will be wrong with this choice of Phase Advance ^"     
  write(mf,'(a)') " ^^^^^^^^^^^^^^^^^^^^^^^^^^^^^^^^^^^^^^^^^^^^^^^^^^^^^^^^^^^^^^^^^^^"
 endif
endif
call print(a_cs,mf); write(6,*);
xs =  a_cs +xs0

 phase=0.0_dp
!!!! finding 
if(np>0) then
flip_phase=-1
 f=>cell%start
  call propagate(xs,state,fibre1=f,fibre2=f0)
  a=xs
  call c_canonise(a,a,a0=a0,phase=phase) 
 beta_0 = (a%v(1).sub.'1')**2 +(a%v(1).sub.'01')**2 
 phase_0=phase(1)
 xs =  a_cs +xs0
 phase=0.0_dp
endif

f=>cell%start
do i=1,cell%n
 call propagate(cell,xs,+state,fibre1=i,fibre2=i+1)

write(mf,*) " end of Magnet ",f%mag%name

 a=xs  ! creates tracked canonical transformation
 xs0=xs ! Save closed orbit at position i
 call c_canonise(a,a_cs,a0=a0,phase=phase) ! A = A_cs o R(phase)
 
 if(np>0) then
  if(f%mag%name=="KICK") flip_phase=-flip_phase
 beta_s = (a_cs%v(1).sub.'1')**2 +(a_cs%v(1).sub.'01')**2 
 phase_s=flip_phase*2*pi*(phase(1)-phase_0)

 dx_co=sqrt(beta_s*beta_0)/2.0_dp/sin(pi*q_0)*cos(phase_s-pi*q_0)


write(mf,'(a)') "Closed orbit as a function of dipole kick strength using formula of Guido "
  if(.not.courant_snyder_teng_edwards) then 
    write(mf,'(a)') " Formula incorrect with this phase advance"
  endif
  write(mf,*) dx_co
write(mf,'(a)') "Closed orbit from TPSA Normal Form "
write(mf,'(a)') "as a function of delta and kick strength "
  call print(a0,mf)
 endif

 write(mf,'(a)') " Phase advance "
 call print(phase(1),mf)
courant_snyder_invariant = r2 * a_cs**(-1)
 write(mf,'(a)') " Invariant "
call print(courant_snyder_invariant,mf)
if(f%mag%name(1:4)=="BEND") then
 betax=courant_snyder_invariant.sub.'02'
 gammax=courant_snyder_invariant.sub.'20'
 alphax=(courant_snyder_invariant.sub.'11')/2.d0

dmu= (atan(gammax*f%next%mag%l-alphax )- atan(-alphax ))/twopi
dmu=dmu+real(phase(1))
 betax=gammax*f%next%mag%l**2-2*alphax*f%next%mag%l + betax
 write(mf,*) " Beta at next element computed ",betax
 write(mf,*) "  Phase at next element ",dmu
endif

xs=xs0+a_cs !updates the real_8 ``xs'' with the ``canonised transformation'' a_cs

f=>f%next
enddo

write(mf,*) " Tune in x from one turn map"
call print(phase_one_turn_map(1),mf)

close(mf)

end program guido_twiss

\end{verbatim}
\pagebreak
\section{MAD-X and Script file  for the lattice in the flat file one_cell.txt}\label{sec:madfiles}

The MAD-X file is a cell of the triple bend achromat of the Advanced Light Source 
 at the

\subsection{The MAD-X file}\label{sec:madfile}

\begin{verbatim}
option,echo,rbarc=false;
L1:drift,L=2.832695;L2:drift,L=0.45698;
L3:drift,L=0.08902;L4:drift,L=0.2155;
L5:drift,L=0.219;L6:drift,L=0.107078;
L7:drift,L=0.105716;L8:drift,L=0.135904;
L9:drift,L=0.2156993;L10:drift,L=0.089084;
L11:drift,L=0.235416;L12:drift,L=0.1245;
L13:drift,L=0.511844;L14:drift,L=0.1788541;
L15:drift,L=0.1788483;L16:drift,L=0.511849;
L17:drift,L=0.1245;L18:drift,L=0.235405;
L19:drift,L=0.089095;L20:drift,L=0.2157007;
L21:drift,L=0.177716;L22:drift,L=0.170981;
L23:drift,L=0.218997;L24:drift,L=0.215503;
L25:drift,L=0.0890187;L26:drift,L=0.45698;
L27:drift,L=2.832696;DS:drift,L=0.1015;
QF1:QUADRUPOLE,L=0.344,K1=2.2474+6.447435260914397E-03;!
QF2:QUADRUPOLE,L=0.344,K1=2.2474;!
QD1:QUADRUPOLE,L=0.187,K1=-2.3368-2.593018157427161E-02;
QD2:QUADRUPOLE,L=0.187,K1=-2.3368;
QFA1:QUADRUPOLE,L=0.448,K1=2.8856;
QFA2:QUADRUPOLE,L=0.448,K1=2.8856;

ksf=(-41.67478927130080+0.3392376315938252)*2.0;
ksd=(56.36083889436033-0.1043679358857811)*2.d0;
sf:sextupole,L=2.0*0.1015,K2=ksf;
sd:sextupole,L=2.0*0.1015,K2=ksd;
sfthin:sextupole,K2=ksf*2.0*0.1015;
sdthin:sextupole,K2=ksd*2.0*0.1015;
VC5:marker;
kick:quadrupole,K1=.0d0;
ALPHA=0.17453292519943295769236907684886;

LBEND=0.86621;

rcav=1.0;
BEND:RBEND,L=LBEND,ANGLE=ALPHA,k1=-0.778741,truerbend=true,ptcrbend=true;
CAV:RFCAVITY,L=0.0,VOLT=-1.0*rcav,LAG=0.0,FREQ=500.0,no_cavity_totalpath;

sfline:line=(ds,sfthin,ds);
sdline:line=(ds,sdthin,ds);
SUP1:LINE=(L1,L2,L3,QF1,VC5,L4,L5,QD1,L6,kick,L7,L8,VC5,BEND,VC5,&
L9,sfline,L10,L11,QFA1,L12,sdline,L13,L14,BEND,L15,L16,sdline,L17,&
QFA2,L18,L19,sfline,L20,BEND,L21,L22,QD2,L23,L24,QF2,L25,L26,VC5,L27);

ALS:line=(sup1);

BEAM,particle=positron,energy=1.5;

use,period=als;
ptc_create_universe;

use,period=ALS;
ptc_create_layout,model=2,method=2,nst=1;!,exact;
ptc_script,file="print_flat_one_cell.txt";
stop;
\end{verbatim}

\subsection{The ptc_script file print_flat_one_cell.txt}\label{sec:scriptfile}

\begin{verbatim}
select layout
1
print new flat file
one_cell.txt
return
\end{verbatim}
\bibliography{biblio.bib}

\begin{thebibliography}{1}

\bibitem{bmada}
D.~Sagan.
\newblock Nucl. \uppercase{I}nstr. and \uppercase{M}eth.
\newblock {\em \bf Phys. Res. A}, 558, 2006.

\bibitem{thebook}
E.~Forest.
\newblock {\em Beam Dynamics: A New Attitude and Framework}.
\newblock Harwood Academic Publishers, Amsterdam, The Netherlands, 1997.

\bibitem{thebook2}
E.~Forest.
\newblock {\em From Tracking Code to Analysis, -Generalised Courant-Snyder
  Theory for any Accelerator Model}.
\newblock (Springer Japan, Tokyo, Japan), 2016.

\bibitem{berz}
M.~Berz.
\newblock Part. \uppercase{A}ccel.
\newblock {\em \bf 24}, 109, 1989.

\bibitem{tpsa}
M.~Berz.
\newblock Nucl. \uppercase{I}nstr. and \uppercase{M}eth.
\newblock {\em \bf A258}, 431, 1987.

\end{thebibliography}
\bibliographystyle{unsrt}








\end{document}